\documentclass[traditabstract,longauth]{aa}
\usepackage{natbib}
\usepackage{epsfig}
\usepackage{txfonts}
\usepackage{graphicx}

\bibpunct{(}{)}{;}{a}{}{,}
\usepackage{subfig}
\usepackage{epsf}
\usepackage{rotating}
\usepackage{graphics}
\usepackage{amssymb}
\newcommand{\corot}{\emph{CoRoT}}

\newcommand{\corotstar}{{Corot-23}}
\newcommand{\corotplanet}{{Corot-23b}}

\newcommand{\Msol}{M$_{\odot}$}
\newcommand{\Rsol}{R$_{\odot}$}
\newcommand{\MJ}{M$_{Jup}$}
\newcommand{\RJ}{R$_{Jup}$}
\newcommand{\ME}{M$_{Earth}$}

\newcommand{\mstar}{{\rm M}$_{\star}$}
\newcommand{\rstar}{{\rm R}$_{\star}$}

\newcommand{\mplanet}{{\rm M}$_{pl}$}
\newcommand{\rplanet}{{\rm R}$_{pl}$}

\newcommand{\kms}{km\,s$^{-1}$}
\newcommand{\ms}{m\,s$^{-1}$}

\newcommand{\vsini}{$v$\,sin\,$i$}

\def\ms{\,m\,s$^{-1}$}         
\def\kms{\,km\,s$^{-1}$}         
\def\vsini{$v$\,sin\,$i$}      
\def\ms{\hbox{\,m\,s$^{-1}$}}         
\def\m2s2{\hbox{\,m$^{2}$\,s$^{-2}$}} 
\def\kms{\hbox{\,km\,s$^{-1}$}}       
\def\vsini{\hbox{$v$\,sin\,$i$}}      
\def\Msun{\hbox{$M_{\odot}$}}             
\def\Rsun{\hbox{$R_{\odot}$}}

\begin{document}
%
%
   \title{Transiting exoplanets from the  \corot\, space mission
   \thanks{The \corot\ space mission, launched on 27 December 2006, has been developed and 
is operated by CNES, with the contribution of Austria, Belgium, Brazil, ESA, Germany, and Spain.  
First \corot\ data are available to the public from the \corot\ archive: http://idoc-corot.ias.u-psud.fr. 
The complementary observations were obtained with MegaPrime/MegaCam, a joint project of 
CFHT and CEA/DAPNIA, at the Canada-France-Hawaii Telescope (CFHT) which is operated by 
NRC in Canada, INSU-CNRS in France, and the University of Hawaii; ESO Telescopes at the La 
Silla and Paranal Observatories under program 184.C0639; the 
OGS telescope operated by the Instituto de Astrof\'\i sica de
Tenerife at Tenerife.}
   }
   \subtitle{XIX. CoRoT-23b: a dense hot Jupiter on an eccentric orbit }
\author{
            Rouan, D.  \inst{1}  
         \and Parviainen, H.  \inst{2,20}
         \and Moutou, C.\inst{4} 
         \and   Deleuil,  M. \inst{4}           
         \and   Fridlund, M.  \inst{5}
         \and A. Ofir  \inst{6}	
         \and Havel, M. \inst{7}
\and Aigrain, S.\inst{8} 
\and Alonso, R.\inst{15}
\and Auvergne, M.\inst{1} 
\and Baglin, A.\inst{1}  
\and Barge, P.\inst{4} 
\and Bonomo, A.\inst{4}
\and Bord\'e, P.\inst{9} 
\and Bouchy, F.\inst{10,11} 
\and Cabrera, J.\inst{3}
\and Cavarroc, C.\inst{9}  
\and Csizmadia, Sz.  \inst{3}
\and Deeg, H.J.\inst{2,20} 
\and Diaz, R.F.\inst{4}
\and Dvorak, R.\inst{12} 
\and Erikson, A.\inst{3}
\and Ferraz-Mello, S.\inst{13} 
\and Gandolfi, D.\inst{5}
\and Gillon, M.\inst{15} 
\and Guillot, T.\inst{7} 
\and Hatzes, A.\inst{14} 
\and H\'ebrard, G.\inst{10,11} 
\and Jorda, L.\inst{4} 
\and L\'eger, A.\inst{9} 
\and Llebaria, A.\inst{4} 
\and Lammer, H.\inst{19} 
\and Lovis, C.\inst{15}
\and Mazeh, T.\inst{6} 
\and Ollivier, M.\inst{9} 
\and P\"atzold, M.\inst{17} 
\and Queloz, D.\inst{15}
\and Rauer, H.\inst{3} 
\and Samuel, B. \inst{1}
\and Santerne, A.\inst{4}  
\and Schneider, J.\inst{16} 
\and Tingley, B.\inst{2,20}
\and Wuchterl, G.\inst{14} 
          }
\institute{
LESIA, UMR 8109 CNRS , Observatoire de Paris, UVSQ, Universit\'e Paris-Diderot, 
5 place J. Janssen, 92195 Meudon, France - daniel.rouan-at-obspm.fr 
\and Instituto de Astrofõsica de Canarias, E-38205 La Laguna, Tenerife, Spain 
\and Institute of Planetary Research, German Aerospace Center, Rutherfordstrasse 2, 12489 Berlin, Germany 
\and Laboratoire d'Astrophysique de Marseille, 38 rue Fr\'ed\'eric Joliot-Curie, 13388 Marseille cedex 13, France 
\and Research and ScientiÞc Support Department, ESTEC/ESA, PO Box 299, 2200 AG Noordwijk, The Netherlands 
\and School of Physics and Astronomy, Raymond and Beverly Sackler Faculty of Exact Sciences, Tel Aviv University, Tel Aviv, Israel 
\and Observatoire de la C\^ote d'Azur, Laboratoire Cassiop\'ee, BP 4229, 06304 Nice Cedex 4, France 
\and Department of Physics, Denys Wilkinson Building Keble Road, Oxford, OX1 3RH 
\and Institut d'Astrophysique Spatiale, Universit\'e Paris XI, F-91405 Orsay, France 
\and Observatoire de Haute Provence, 04670 Saint Michel l'Observatoire, France 
\and Institut d'Astrophysique de Paris, 98bis boulevard Arago, 75014 Paris, France 
\and University of Vienna, Institute of Astronomy, T\"urkenschanzstr. 17, A-1180 Vienna, Austria  
\and IAG, Universidade de Sao Paulo, Brazil  
\and Th\"uringer Landessternwarte, Sternwarte 5, Tautenburg 5, D-07778 Tautenburg, Germany 
\and Observatoire de l'Universit\'e de Gen\`eve, 51 chemin des Maillettes, 1290 Sauverny, Switzerland 
\and LUTH, Observatoire de Paris, CNRS, Universit\'e Paris Diderot; 5 place Jules Janssen, 92195 Meudon, France 
\and Rheinisches Institut f\"ur Umweltforschung an der Universit\"at zu K\"oln, Aachener Strasse 209, 50931, Germany 
\and University of Li\`ege, All\'ee du 6 ao\^ut 17, Sart Tilman, Li\`ege 1, Belgium 
\and Space Research Institute, Austrian Academy of Science, Schmiedlstr. 6, A-8042 Graz, Austria  
\and Universidad de La Laguna, Dept. de Astrof\'\i sica, E-38200 La Laguna, Tenerife, Spain 
}

   \date{v5 -  Received August 23, 2011; accepted  October 4, 2011}

 \abstract{We report the detection of CoRoT-23b, a  hot Jupiter transiting in front of its host star with a period of 3.6314 $\pm$    0.0001 days. 
This planet was discovered thanks to photometric data secured with the CoRoT satellite, combined with spectroscopic radial velocity  (RV) measurements.  
   A photometric search for  possible background eclipsing binaries  conducted at CFHT and OGS  concluded with a very low risk   of false positives. 
The usual techniques of combining  RV and transit  data simultaneously  were used to derive stellar and planetary parameters.  
   The planet has a mass of Mp = 2.8  $\pm$ 0.3 
\MJ\,,   a radius of \rplanet = 1.05  $\pm$  0.13 
\RJ\, , a density of $\approx$ 3 g cm$^{-3}$. RV data also clearly reveal a non zero eccentricity of e = 0.16 $\pm$ 0.02.  The planet orbits  a   \textit{mature}  G0 main sequence star of V =15.5 mag,  with a mass \mstar = 1.14  $\pm$ 0.08 
 \Msol , a radius \rstar  = 1. 61  $\pm$  0.18 
 \Rsol\, and quasi-solar abundances.
   The age of the system is evaluated to be 7 Gyr, not far from the transition to subgiant, in agreement with the rather large stellar radius. 
  The two features of a significant eccentricity of the orbit  and of a fairly high  density  are
   fairly uncommon for a hot Jupiter. The high density is, however, consistent with  a model of contraction of a planet at this 
   mass, given  the age of the system.  On the 
   other hand, at  such an age, circularization   is  expected to  be  completed. In fact, we show that for this planetary 
   mass and orbital distance, any initial  eccentricity should  not totally vanish after 7 Gyr, as long as the  tidal quality factor Q$_{p}$ is    more than a few 10$^5$, a value that is  the lower bound of the usually expected range. Even if \corotplanet\, features a density and an eccentricity that are atypical of a hot Jupiter, it is thus not an enigmatic object.
  }
 \authorrunning{D. Rouan  et al.}
  \titlerunning{\corotplanet\,   \corotplanet\, an eccentric and dense  hot Jupiter}
   \maketitle
%
\section{Introduction}
 
  Transiting exoplanets are especially interesting in the study of 
exoplanet structure and evolution, since a transit allows  measurement of the radius of the 
exoplanet and also ensures that the inclination is close to 90$^{o}$,  therefore that the true 
mass is the minimum mass derived from  RV. Transiting planets thus provide 
strong constraints on planetary internal structures and evolution models by yielding mass, 
radius, and density.  Since H209458b \citep{2000ApJ...529L..45C}, the first transiting planet, extensive ground-based 
photometric surveys (OGLE, TrES, XO, HAT and SuperWASP) and, more recently, space 
missions (MOST, CoRoT, Kepler) provide many transiting planetary candidates, and today (June 2011) 172 exoplanets have been confirmed with well constrained masses for a large majority of cases.

  Since the beginning of 2007, the space mission \corot\ performs wide-field stellar photometry at ultra-high precision from space \citep{1998EM&P...81...79R,2006cosp...36.3749B}. Currently, during an
   observing run up to 6,000 stars\footnote{12,000 until Oct 2009, but at this date,  control of one of the two CCDs was lost, probably because of a high energy particle hit on the onboard computer.} can be monitored simultaneously and continuously over  periods of 20 to 150  days of  observation.  \corot\, is thus particularly well-suited to detect planets with rather short orbital periods, from less than one day to 50 days, and sometimes more \citep{2010Natur.464..384D}. \corot\, has already detected 22 planets or brown dwarves in June 2011, and  in this paper, we report the discovery of  a 23 rd transiting massive planet detected around the main sequence  G0 star \corotstar.   

As in all  transit surveys, ground-based follow-up is mandatory for 
 confirming a transiting planet candidate. In the case of \corotplanet, the one planet whose
 detection is claimed hereafter, a standard follow-up 
program was performed, including photometry, spectroscopy, and  
 RV measurements, using different ground-based facilities over the world. 
 
We present  the photometric analysis of the  \corot\,  data from which we discovered  this transit 
candidate in Sect. \ref{sec:corotLC},  while the photometric 
follow-up   and results of the RV measurements that allowed its planetary nature to  be secured, are
depicted in in Sects. \ref{sec:ground_photom} and  \ref{sec:RV}, respectively. 
The stellar parameters  derived from spectroscopic analysis are  presented in
Sect.  \ref{sec:stellar_params}, while  final planetary and stellar parameters are given in Sect.  
\ref{sec:stellar_planet_param}.   Several questions about the 
 eccentricity and the density  of the planet, in relation with its age, are discussed in Sect.  \ref{sec:discussion}.

\section{CoRoT observations}
\label{sec:corotLC}

   \begin{figure}
   \centering
\includegraphics[width=8cm]{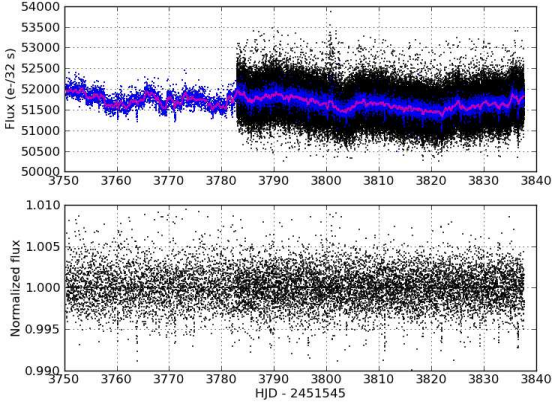}
   \caption{ CoRoT LC of the target \corotstar. The apparent change of regime at CoRoT JD 3783 corresponds to the change in the sampling period from   512s to  32s.   Upper panel: raw data (in black), at a unique sampling period of 512 s (blue) and after a median-smoothing on 10 h (purple). Lower panel : LC after applying a detrending filter.  
             \label{fig:Corot-23b_LC}}%
    \end{figure}

\begin{figure}[!th]
\begin{center}
\includegraphics[width=8.5cm]{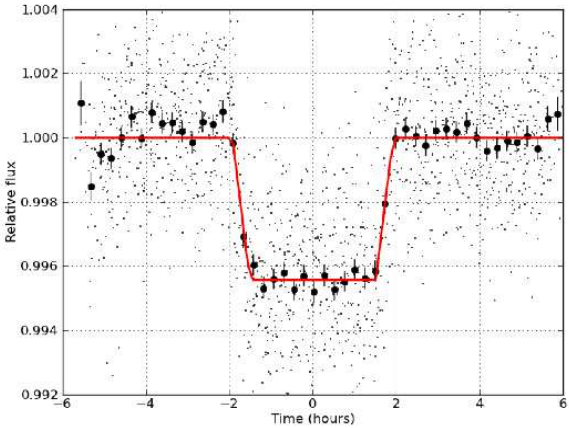}
\caption{Phase-folded LC  of \corotplanet\, using the ephemeris given in Table~\ref{starplanet_param_table} with, superimposed, a simple trapezoidal fit. } 
\label{fig:corot-23b-LC0-folded}
\end{center}
\end{figure}


The star \corotstar\, was observed during the LRc05  long run of  \corot\, (02/04/2010 to 05/07/2010) towards the Serpens Cauda constellation. Its ID is given in Table~\ref{StarID}, based on the {\sl Exo-Dat} database 
\citep{2009AJ....138..649D}.
Spurious spikes and stellar variations at frequencies outside the range expected for planetary transits were removed with low- and high-pass filters, as illustrated in Fig.\ref{fig:Corot-23b_LC}. Then, different detection algorithms were used and 24 individual transits were eventually detected.

\begin{table}[h]
\caption{ IDs, coordinates and magnitudes.}            
\centering        
\begin{minipage}[!]{7.0cm}  
\renewcommand{\footnoterule}{}     
\begin{tabular}{lcc}       
\hline\hline                 
CoRoT window ID &  LRc05\_E2\_4607 \\
CoRoT ID & 105228856 \\
USNO-A2 ID  & 0900-13361093 \\
2MASS ID   &  18390782+0421281 \\
\\
\multicolumn{2}{l}{Coordinates} \\
\hline            
RA (J2000)  & 279.782615  \\
Dec (J2000) & +4.35780  \\
\\
\multicolumn{3}{l}{Magnitudes} \\
\hline
\centering
Filter & Mag & Error \\
\hline
B$^a$  & 16.96 & $\pm$ 0.23 \\
V$^a$  & 15.63 & $\pm$  0.07\\
r'$^a$ & 15.038 & $\pm$  0.043\\
i'$^a$ & 14.198 &$\pm$ 0.034 \\
J$^b$  &12.94  & $\pm$ 0.02  \\
H$^b$  &12.45&  $\pm$ 0.02  \\
K$^b$  &12.36 & $\pm$ 0.02   \\
\hline\hline
\vspace{-0.5cm}
\footnotetext[1]{Provided by Exo-Dat (Deleuil et al, 2008);}
\footnotetext[2]{from 2MASS catalog.}
\end{tabular}
\end{minipage}
\label{StarID}      
\end{table}

 
 The main transit parameters, i.e., period, central date, ingress/
egress duration, total duration, and relative depth, are estimated using the  differential evolution global optimization method  \citep{2009ApJ...690L...1V}.  The model is parameterized by the squared  star-planet radius ratio, transit center, period,  inverse of the transit width 
\citep{2010MNRAS.407..301K},  squared impact parameter, limb-darkening cofficient(s),  a zeropoint correction, and the eccentricity, when provided. The MCMC simulations also include a scale factor
to the point-to-point scatter estimated from the lightcurve as a free
parameter (somewhat similar to the way the errors are handled by
\citeauthor{2007MNRAS.381.1607G} \citeyear{2007MNRAS.381.1607G} ), so these
distributions are marginalized over both the uncertain zeropoint and
point-to-point scatter in order to obtain reliable 
results.

 The different  analysis agree on a  period of 3.6313 $\pm$  0.0001  days.  Figure \ref{fig:corot-23b-LC0-folded}, where all 
transits are summed using this period, shows the clear  transit signal and its fit by a trapezoid. We evaluated   $\tau_{23}$ = 3.383 h and $\tau_{14}$= 3.888 h for   the inner and outer durations, respectively and  $\Delta F/F$ =   4.3 10$^{-3}$ $\pm$  10$^{-4}$ for the depth of the transit.  Those values take an estimate of the contamination of 7.2\% by stars in the field into account, based on modeled CoRoT point spread function (\emph{PSF}). 
A more complete analysis was performed in a second step, once the RV data analysis was able
to provide an eccentricity estimate (see sect. \ref{sec:RV}). 
 
\section{Ground-based observations}
\label{sec:ground_FU}

\subsection{ Photometric time series }
\label{sec:ground_photom}


Whenever  clear periodic transit-like events are detected in a \corot\ LC and when the candidate survives the set of tests 
performed to rule out obvious stellar systems  \citep[see]{2009A&A...506..491C}, a ground-based 
follow-up program is initiated.  The goal is to check further for possible contaminating  eclipsing 
binaries (CEBs) whose  \emph{PSF} could fall within the \corot\ photometric 
mask. This is done by  searching for photometric variations on nearby stars 
during the predicted  transits.

The  \corot\,  exoplanet channel has a large \emph{PSF} extending over  a roughly ellispoidal area of 60 arcsec $\times$ 32 arcsec, which implies a significant probability that candidates detected 
in the \corot\, data arise from nearby CEBs. The photometric follow-
up program of  \corot\, candidates intends to identify such CEBs, comparing observations during 
predicted transit-times with observations out of transit. Time-series follow-up  is described in more detail in \citet{2009A&A...506..343D}. 

\begin{figure}
   \centering
\includegraphics[width=8cm]{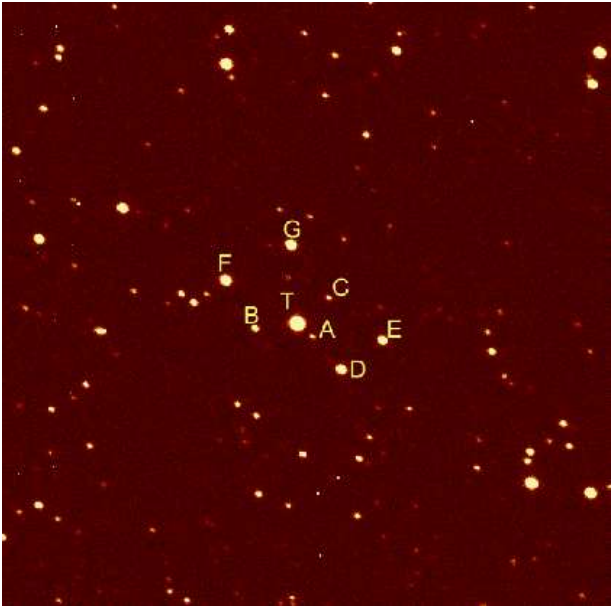}
   \caption{    The field of \corotstar , as extracted from a MEGACAM frame of 10 sec
   taken at CFHT. The potentially contaminating stars are labeled   with letters from A to G. 
   The size of the field is $\approx 3 \times 3 \, arcmin^{2}$} \label{fig:corot-23b-cfht} %
    \end{figure}

Figure \ref{fig:corot-23b-cfht} shows that the field is rather crowded around the target star (labeled with a  T). 
For any  nearby star around the  \corotstar\,  target, we calculated the expected 
eclipse amplitude \emph{ if} this star was the source of the observed dips. Calculation of this 
amplitude is based on a model of the stellar \emph{PSF}, the shape of the photometric aperture, 
and the position and magnitudes of the target and the contaminating stars, respectively.  
These amplitudes were found for the stars labeled
A,B,C,D, and G in Fig. 3 to be in the range of 0.2 - 0.9 mag, whereas
all other stars are either too far away or too faint to be of concern.

\begin{figure}
   \centering
\includegraphics[width=8cm]{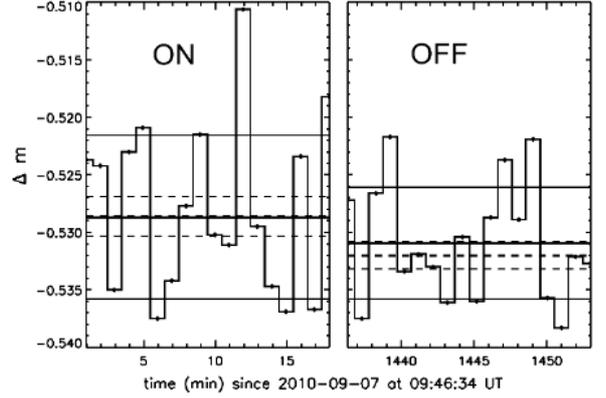}
   \caption{     Relative magnitude variation of \corotstar , as measured with CFHT-MEGACM, at two time
    intervals of the night 07 Sept. 2010, the first one (noted ON) during the predicted transit.     
   The expected  variation
   due to the transit is detected with a significant signal-to-noise ratio as shown by the 1 $\sigma$ range (dotted lines) for each sequence; it corresponds in amplitude and sign  to the 
   predicted one.\label{fig:corot-23b-photFU-cfht}}%
    \end{figure}

Observations to identify whether any of the concerned stars show such
amplitudes were then done on two telescopes:  the 1m OGS on Izaña,
Tenerife, and the CFHT, Hawaii. The observations on OGS were taken during a
transit on 13 Aug. 2010, and the off-transit comparison on 10 Aug.
2011. The CFHT observations with MEGACAM \citep{2003SPIE.4841...72B} were
performed during  a transit on 7 Sept. 2010 and one day after, at the same hour of
the night ( 20 exposures of 10 s integration  during 16 minutes).  From both data
sets, photometry was extracted through classical differential
photometric techniques and the stars on-and off-transit brightness
were compared. In no case did the concerned stars show any relevant
brightness variations. The times of the transit events in Aug. resp.
Sept. 2010 could be predicted with an error of about 30 minutes,
and the observations were therfore performed during on-transit  with
high probability, so any deep eclipses on these stars can be
excluded. The CFHT observations futhermore detected a marginally
significant signal on the target (see Fig. \ref{fig:corot-23b-photFU-cfht}), with an ampltiude of
0.6$ \pm$ 0.2\%, compatible with the 0.38\% amplitude in CoRoT's data. We
concluded that there is no CEB in the field that can explain the
CoRoT periodic signal and that the target star should be a good
candidate for harboring a transiting planet. 

\subsection{Radial velocity }
\label{sec:RV}

\begin{table}[h]
\begin{center}{
\caption{HARPS RV observations of \corotstar.}
\begin{tabular}{llll}
\hline
Jul Date   &   RV&     error& bis\\
-2400000. &km/s &km/s& km/s\\
\hline
55657.87089 &-35.419 &0.0638 &0.0519\\
55681.91406 &-34.862 &0.0743 &0.0814\\
55683.85633 &-35.521 &0.0763 &0.0989\\
55684.77013 &-35.199 &0.0809 &0.0149\\
55685.76995 &-34.758 &0.0558 &-0.0449\\
55686.78087 &-35.271 &0.0777 &-0.0009\\
55687.80642 &-35.455 &0.0804 &0.0716\\
55689.82207 &-34.882 &0.0830 &0.4105\\
55695.91662 &-35.168 &0.0530 &-0.0729\\
55716.84835 &-35.620 &0.1632 & 0.2432\\
\hline
\label{Tab-HARPS}
\end{tabular}}
\end{center}
\end{table}

\begin{figure}
\begin{center}{
\includegraphics[width=8.5cm]{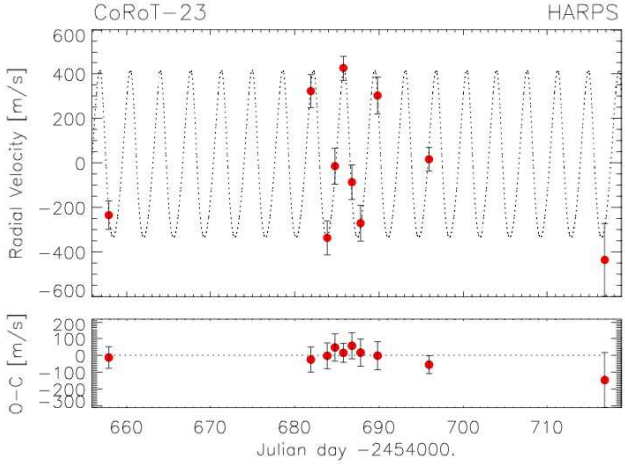}
\caption{HARPS RV measurements of \corotstar\, obtained in  2011, versus 
time. The data points are shown with 1$\sigma$ error bars.  Superimposed is a Keplerian 
orbital curve of a 3.6313  day period planet at the CoRoT ephemeris. We conclude that the planet has a mass \mplanet $\approx$ 2.8 $\pm$ 0.3  \MJ.
}
\label{fig:RV-HARPS-time}}
\end{center}
\end{figure}

\begin{figure}
\begin{center}{
\includegraphics[width=8.5cm]{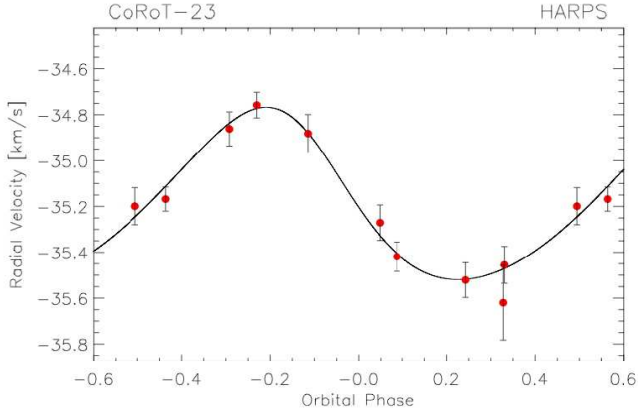}
\caption{Phase-folded HARPS RV measurements of \corotstar\, showing the clear eccentricity of  0.16 when fitted by a Keplerian elliptical orbit.
}
\label{fig:RV-HARPS-folded}}
\end{center}
\end{figure}

\begin{figure}
\begin{center}{
\includegraphics[width=8.5cm]{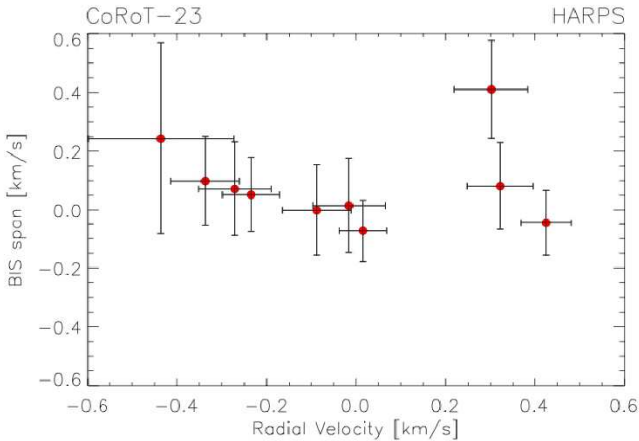}
\caption{Bisector span vs radial velocity of the HARPS RV measurements of \corotstar .
}
\label{fig:RV-HARPS-bis}}
\end{center}
\end{figure}

Nine spectra were gathered with HARPS on the 3.6m telescope at La Silla Observatory (ESO, Chile), from 5 April from 13 May 2011. Their signal-to-noise ratio varies from six to ten at 550 nm in one-hour exposures. The spectra were extracted by the HARPS pipeline, and the RV was derived from a cross-correlation with a G2 mask \citep{1996A&AS..119..373B,2002A&A...388..632P}. The cross-correlation function shows a single peak of width 12.9 km/s, indicating a slightly rotating star. The average error bar on the velocity is 72 m/s, while the velocity has a variation of 760 m/s peak-to-peak with an $rms$ of 280 m/s. 
The HARPS measurements versus time are plotted in Fig. \ref{fig:RV-HARPS-time}.  When fitted at the CoRoT ephemeris, using AMOEBA convergence and bootstrap analysis of the errors, the RV signal is compatible with a slightly eccentric Keplerian orbit with a semi-amplitude of 377  $\pm$ 11 m/s and   {\rm e} = 0.16 $\pm$ 0.02, as illustrated on Fig. \ref{fig:RV-HARPS-folded}. 

As indicated by Fig. \ref{fig:RV-HARPS-bis},  the bisector span does not show any significant correlation with the velocity, a good indication that the velocity signal comes from a planetary companion rather than produced by stellar activity or by a background eclipsing binary.    

We conclude that the star CoRoT-23 hosts a transiting planetary companion of 2.8 $\pm$ 0.3 Jupiter masses at a semi-major axis of 0.048 AU on a significantly eccentric orbit. 

\subsection{Spectroscopy and stellar parameters}
\label{sec:stellar_params}

\begin{figure}
\begin{center}{
\includegraphics[width=8.5cm]{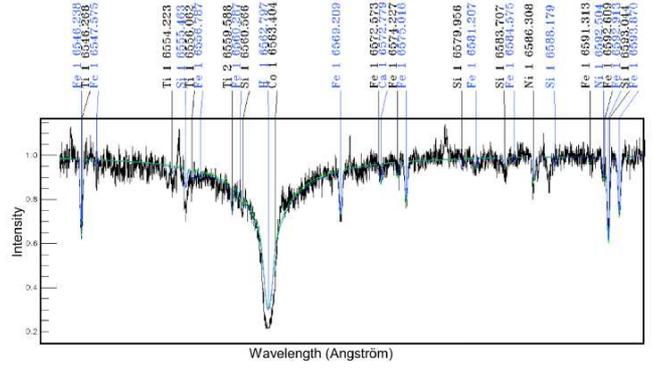}
\caption{  Part of the HARPS   spectrum centered on the H$_{\alpha}$ line.
A fit of the spectrum using synthetic stellar spectra of Kurucz (http://kurucz.harvard.edu/grids.html) is superimposed with
identification of several lines.
}
\label{fig: corot-23b-spectrum}}
\end{center}
\end{figure}

The central star was spectroscopically analyzed using the HARPS data set. 
Figure \ref{fig: corot-23b-spectrum} shows a significant sample of the HARPS spectrum used for this analysis. By 
comparing the spectra  with a grid of stellar templates, as described in \citet{2003A&A...405..149F} 
and \citet{2008ApJ...687.1303G}, or using SME \citep{1996A&AS..118..595V}, we derived the spectral type and luminosity class 
of the star and the stellar parameters, which are summarized in  Table \ref{starplanet_param_table}. In short, the star is a  G0 V with Teff = 5900 $\pm$ 100 K, \mstar = 1.14 $\pm$ 0.08 \Msun, logg = 4.3 $\pm$ 0.2  and  features solar type abundances : [Fe/H]:   [ 0.05 , 0.10]. 
Given the stellar
density deduced from the transit data, the radius of the star was evaluated is 1.61  $\pm$  .18 \Rsun. 
The stellar rotation period is estimated to be 9.2 $\pm$ 1.5 days, as  derived from  V $\sin i$ = 8 $\pm$ 1 km/s. This value could match a peak at a period of 10.5   days on the periodogram of the \corot\, light curve.

The  age of the star has been evaluated to be between 6 and 9  Gyr, using evolution tracks of main sequence and post main sequence stars in the T$_{eff}$ vs M$^{1/3}$/R diagram.  The possibility that \corotstar\, could be a pre-MS star is excluded since an unlikely age of 18 Myr would be constrained. Given the rather low density
derived from  the complete analysis and the large corresponding radius of the star (1.61 \Rsun), it
is  likely that the star is at an evolution stage close to leaving the main sequence  and  evolving to a subgiant.    The star/planet co-evolution diagram of Fig. \ref{fig:C23_co-evolution} (see Sect.\ref{sec:discussion})  also points to an age of more than 5 Gyr.

{\bf 

\begin{table*}
\vspace{1cm}
\centering
\caption{Planet and star parameters.}            
\vspace{1cm}
\begin{minipage}[t]{13.0cm} 
\setlength{\tabcolsep}{10.0mm}
\renewcommand{\footnoterule}{}                          
\begin{tabular}{l l}        
\hline\hline                 
\\
\multicolumn{2}{l}{\emph{Ephemeris}} \\
\hline
Planet orbital period $P$ [days] & 3.6313 $\pm$  0.0001 \\
Primary transit epoch $T_{tr}$ [HJD-2400000]  & 55308.939  $\pm$ 0.001  \\
Primary transit duration $d_{tr}$ [h] & 3.888 $\pm$ 0.054 \\
& \\
\\
\multicolumn{2}{l}{\emph{Results from radial velocity observations}} \\
\hline    

Orbital eccentricity $e$  & 0.16  $\pm$ 0.02 \\
Argument of periastron $\omega$ [deg] & 52 $\pm$ 9 \\ 
Radial velocity semi-amplitude $K$ [\kms] & 0.377 $\pm$  0.034\\
Systemic velocity  $V_{r}$ [\kms] & -35.182 $\pm$ 0.003 \\
O-C residuals [\ms] &  35   \\
& \\
\\
\multicolumn{2}{l}{\emph{Fitted transit parameters}} \\
\hline
Radius ratio $k=R_{p}/R_{*}$ & 0.0671  $\pm$ 0.0010  \\
Linear limb darkening coefficient~$^a$ $u$ & 0.33  $\pm$ 0.08 \\
Impact parameter$^c$ $b$ & 0.56  $+$0.10 $-$0.23 \\
& \\
\\
\multicolumn{2}{l}{\emph{Deduced transit parameters}} \\
\hline
Scaled semi-major axis $a/R_{*}$~$^b$ & 6.85 $\pm$ 0.60 \\
$M^{1/3}_{*}/R_{*}$ [solar units]& 0.641 $\pm$ 0.065 \\
Stellar density $\rho_{*}$ [$g\;cm^{-3}$] & 0.50 $\pm$ 0.15 \\
Inclination $i$ [deg] & 85.7 +2.6 -1.5   \\
& \\
\\
\multicolumn{2}{l}{\emph{Spectroscopic parameters }} \\
\hline
Effective temperature $T_{eff}$[K] & 5900 $\pm$ 100\\
Surface gravity log\,$g$ [dex]& 4.3 $\pm$ 0.2 \\
Metallicity $[\rm{Fe/H}]$ [dex]& 0.05 $\pm$ 0.1 \\
Stellar rotational velocity {\vsini} [\kms] &  9.0 $\pm$ 1 \\
Spectral type &  G0 V \\
& \\
\\
\multicolumn{2}{l}{\emph{Stellar and planetary physical parameters from combined analysis}} \\
\hline
Star mass [\Msun] &  1.14 $\pm$ 0.08 \\
Star radius [\Rsun] & 1.61  $\pm$  .18  \\
Distance of the system [pc] &  600 $\pm$  50 \\
Stellar rotation period $P_{rot}$ [days]  &   9.2  $\pm$  1.5  \\
Age of the star $t$ [Gyr] & 7.2   -1 +1.5    \\
Orbital semi-major axis $a$ [AU] &  0.048 $\pm$  0.004 \\
Planet mass $M_{p}$ [M$_J$ ]$^d$ &  2.8  $\pm$ 0.3 \\
Planet radius $R_{p}$[R$_J$]$^d$  & 1.05 $\pm$  0.13  \\
Planet density $\rho_{p}$ [$g\;cm^{-3}$] &   3.0 $\pm$  1.1  \\
Equilibrium temperature $^f$  $T^{per}_{eq}$ [K] &  1660 \\
\\
\hline       

\vspace{-0.5cm}
\end{tabular}
\end{minipage}
\label{starplanet_param_table}  
    
\end{table*}

}
 Regarding  the distance estimate we have first converted the 2MASS magnitudes into
the SAAO system with the relations of \cite{2001AJ....121.2851C}.
The calculated colors J-H and H-K are 0.49 $\pm$ 0.03 and
0.09 $\pm$ 0.03, respectively.
Using \cite{2001ApJ...558..309D} for the intrinsic colors of
a G0 star, we derive the  color excesses from [B-V] to [H-K] and, using 
an extinction law towards the Galactic center region \citep{2003MNRAS.338..253D},  A$_{K}$ = 0.18 and A$_{V}$ = 2.42. 
Taking for the spectral type of \corotstar\, an absolute magnitude of
$M_V= 4.4  \pm 0.1$ \citep{1981Ap&SS..80..353S}, the derived distance, based either on (B-V)
or (J-K) indices, is   550 and 600 pc, respectively. We assume  that the infrared derivation is somewhat more robust, so  we retain it in Table \ref{starplanet_param_table}.
\\
\\
    
\section{Final stellar and planetary parameters }
\label{sec:stellar_planet_param}

\begin{figure}
   \centering
\includegraphics[width=9cm]{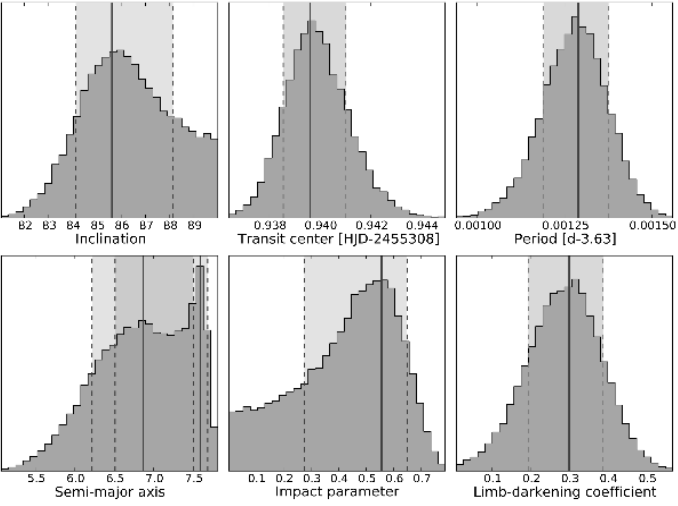}   
   \caption{Probability distribution of six transit parameters,   obtained from the differential evolution global optimization  method applied to the \corot\, transit signal,  assuming a linear limb-darkening law and taking the eccentricity into account.
              \label{fig:C23b-histo} 
              }%
    \end{figure}

Once the stellar parameters are determined and the radial velocity analysis done, a
more consistent and more accurate analysis of the \corot\, LC can be performed in a second step. 
Taking the eccentricity into account as measured thanks to RV data analysis (see Sect. \ref{sec:RV}), and the stellar parameters, as derived in Sect. \ref{sec:stellar_params} we computed the complete solution for the different parameters, summarized in  Table \ref{starplanet_param_table}.  The uncertainty on each parameter is derived from its statistical distribution, as  shown in Fig. \ref{fig:C23b-histo} for six of them.

To  summarize the main characteristics of the planet in one sentence: \corotplanet\, is a dense hot Jupiter ($R_{p}$ = 1.05 R$_J$, $M_{p}$ =  2.8 M$_J$, $\rho_{p}$ =   3.0 $g\;cm^{-3}$, $a$ =  0.048 AU) with a significant eccentricity ($e$ = 0.16) orbiting a mature  G0 star.   


\section{Discussion}
\label{sec:discussion}

  \begin{figure}
   \centering
\includegraphics[width=8cm]{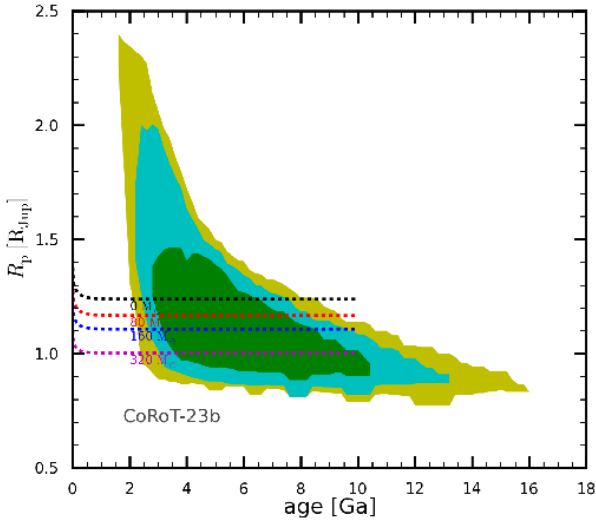}   
   \caption{Diagram describing the probability that \corotplanet\, has a given age and radius (the green area  corresponds the highest probability)
, given the set of parameters deduced from the observations (photometry, spectroscopy, and radial velocity) and using a combination of stellar and planetary evolution codes to constrain the probability. The four dotted lines describe the evolution tracks of a giant gaseous planet with the different core masses as labeled.               \label{fig:C23_co-evolution} 
              }%
    \end{figure}

\begin{figure}
   \centering
\includegraphics[width=8cm]{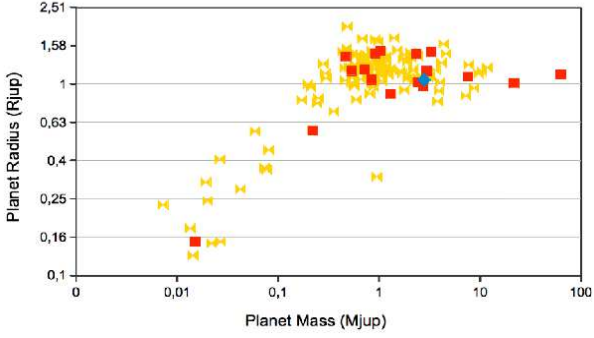}   
   \caption{  Comparison of the different transit planets  discovered by \corot\, (red squares) or by 
   other experiments (yellow bowties), as well as \corotplanet\, (blue diamond) in a mass/radius 
   diagram. \corotplanet\,   lies on the curve that joins the densest  \corot\, planets.
              \label{fig:density} 
              }
    \end{figure}

\corotplanet \, appears as a hot Jupiter that is both rather dense and features a significant eccentricity, two characteritics that are  not typical for this class of exoplanets. 
On the mass/radius diagram of  Figure \ref{fig:density} we plotted the various transit planets   discovered by \corot\, or by other experiments as well as \corotplanet. \corotplanet\, appears  to lie  at  the lower boundary of the domain of
planets discovered by \corot, meaning that it is among the densest members. Does this property translate the age of the system, which is  close to 7.5 Gyr, a time sufficient for the almost full contraction of the planet? This is one plausible explanation, that is consistent with a
combined modeling of the star 
\citep{2008Ap&SS.316...61M, 2008A&A...482..883M} 
 and the
planet  \citep{1995A&AS..109..109G}
 in a system \citep{2011A&A...527A..20G, 2011A&A...531A...3H} 
 , as illustrated in Fig. \ref{fig:C23_co-evolution} where  the  probability that \corotplanet\, has a given age and radius is indicated by a color, the green area  corresponding to the highest probability.  Evolution tracks of a giant gaseous planet with different core masses  are superimposed, and  suggest that a rather massive core (300 \ME) is required, if the age of the system is indeed close to 7.5 Gyr. 
 However, the observations do not constrain the radius of the planet very well (13\% of uncertainty), and thus other solutions for the core mass remain, ranging from 0 to about 350-400 MEarth.
In our planetary evolution calculations, we assumed that (i) all heavy elements are grouped into a dense central core, surrounded by a solar-composition H/He envelope 
 \citep[see ][for a discussion]{2008A&A...482..315B}; 
 (ii) 0.25\% of the incoming stellar flux is dissipated deep into the planetary interior \citep{2008PhST..130a4023G}    
We estimate that these assumptions would have a
small effect on the determination of the core mass (less than 10\%) as compared to the uncertainty due to the observations.

\begin{figure}
   \centering
\includegraphics[width=8cm]{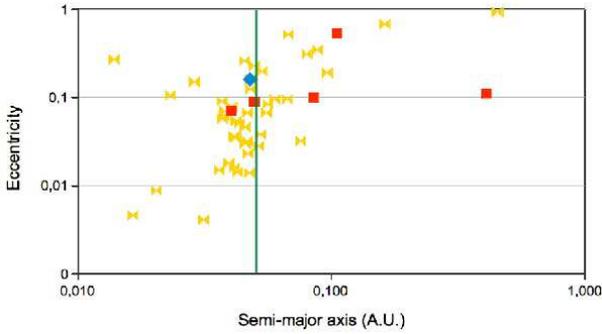}   
   \caption{  Diagram of eccentricity vs semi-major axis  of the different transiting planets   that have no-zero eccentricity. We display with different symbols the ones discovered by \corot\, (red squares), by other experiments (yellow bowties), and by \corotplanet\, (blue diamond). The vertical line marks an arbitrary frontier between hot and warm Jupiters. 
              \label{fig:eccentricity} 
              }
    \end{figure}

 As regards the eccentricity, \corotplanet, with $e$ = 0.16, has the fourth most eccentric orbits of all planets discovered by \corot and the 14th of the 172 confirmed transit planets (June 2011);  it features one of the most eccentric orbits of the  hot Jupiter class, as illustrated in Fig. \ref{fig:eccentricity}. 
 For instance, if we consider a boundary at a semi-major axis of 0.05 A.U. between hot and warm Jupiters,  the median eccentricity (excluding all zero eccentricity planets) is 0.06 for the former (closer orbits) and 0.11 for the later (wider orbits). \corotplanet\, is clearly  distinct from other hot Jupiters of its group in this respect. It is generally admitted and statistically consistent that the low eccentricity of hot Jupiters results from circularization through tidal interaction between the star and the planet, more precisely by tides raised on the planet  and/or during  the  migration,  since the disk also tends  to damp the eccentricity   \citep{2008DDA....39.1515M}. The last mechanism  however has a much shorter time constant (0.1-10 Myr)  than the age of the system. On the other hand, in a few recent cases, no-zero eccentricities of planets on close orbits \citep{2010ApJ...723L..60H,2010ApJ...708..224H, 2009ApJ...693.1920H} raised  interest in alternate scenarii based on the presence of a companion that gravitationnaly perturbes the orbit.  Is that the case here? 
 This being said, we  note that the circularization efficiency  and time constant depend on several parameters, such as the mass of the planet and the orbital distance. \cite{2011MNRAS.414.1278P} 
 examine this question from a statistical point of view and show that the frontier between fully 
 circularized planets and eccentric ones is fairly well defined in the diagram of mass vs orbital 
 distance  (see their Fig. 3). Not all short-period planets have zero eccentricity, even in a tide-driven 
 circularization scenario.
  
  In these conditions, does that no-zero eccentricity of  \corotplanet\,  result from a normal secular 
  evolution  or does it reflect  the recent gravitational effect from other putative planets in the system 
  is a question that   deserves some further studies. 
  
\begin{figure}
   \centering
\includegraphics[width=8cm]{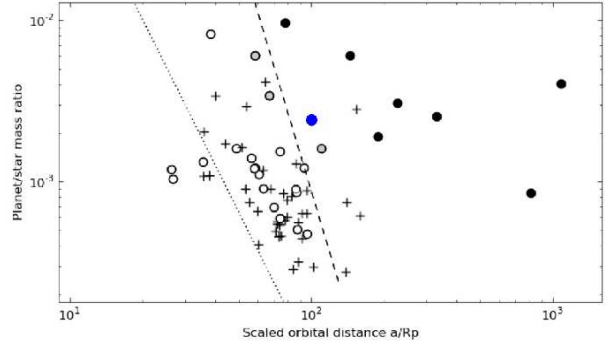}   
   \caption{  \corotplanet\, (blue dot) added on the  M$_{pl}$/ \mstar vs a / R$_{pl}$ diagram proposed by \cite{2011MNRAS.414.1278P}.  \corotplanet\, appears to lie outside the band of circularized planets.
              \label{fig:c23b-mpVsa} 
              }
    \end{figure}

  We first added the case of \corotplanet\, in Fig. 3 of \cite{2011MNRAS.414.1278P}, i.e.  the M$_{pl}$/ \mstar vs $a$/M$_{pl}$ plane, as shown in Fig. \ref{fig:c23b-mpVsa}. \corotplanet\, appears not far, but clearly outside the band of circularized planets. This is a first hint that a perturbating companion is not  mandatory and that the no-zero eccentricity of \corotplanet\, could just be due to a long time constant for circularization because of its mass and distance to the star. 
  
\begin{figure}
   \centering
\includegraphics[width=8cm]{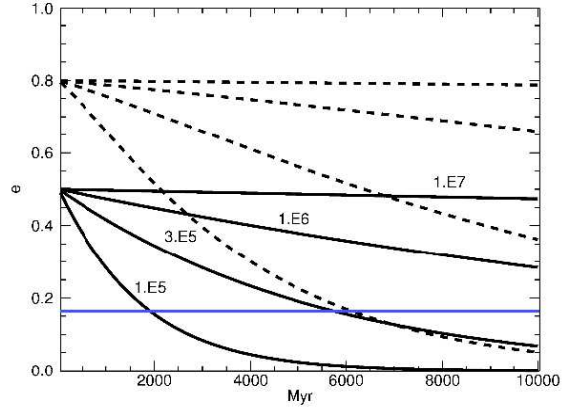}   
   \caption{   Evolution of the eccentricity of \corotplanet\, with time (in Myr) for two initial conditions (e = 0.5: solid lines and e = 0.8: dash lines) and four different values of the tidal quality factor Q$_{p}$: 10$^5$, 3. 10$^5$, 10$^6$, and 10$^7$. The evolution of e was computed using Eq. 17 of \cite{2010ApJ...725.1995M}  and the parameters of Table \ref{starplanet_param_table}.  The horizontal blue line corresponds to the measured eccentricity of \corotplanet.
              \label{fig:C23-damping}
              }
    \end{figure}

We  tried then to put some constraints on the  tidal quality factor Q$_{p}$ that governs the damping of the eccentricity of 
\corotplanet\,. Following \cite{2010ApJ...725.1995M}, we assume that the eccentricity evolution
 only depends  on the tidal 
dissipation inside the planet, so that $d{\rm e}/d{\rm t}$ is described by Eq. 17 of 
\cite{2010ApJ...725.1995M}. We  
computed the evolution of $e$ with time starting from two situations of a rather high eccentricity (0.8
 and 0.5) just after the 
formation of the planet. This is illustrated in Fig. \ref{fig:C23-damping} where several acceptable
 values for  Q$_{p}$ were 
considered. We find that the damping   would be fast enough to circularize the orbit only for the lowest initial
 eccentricity and the lowest value of Q$_{p}$.   
 We conclude that, if the \corotstar\, system is indeed mature with an age,  say over 5 Gyr,
  then the damping  of the 
 eccentricity should have been slow, a condition reached as soon as  Q$_{p}$ is greater that 
 3.\,10$^{5}$.  This value appears to 
 be well within the pausible range and even at the lower bound: no unusual response of  
  \corotplanet\, to tides would then be 
 required to explain its  eccentricity.
 
We note also that, compared to Corot-20 \citep{2011A&A..Deleuil},   the system  has a larger stellar 
property factor  \citep[as defined by][]{2002ApJ...568L.117P} by a factor of 10  and a larger Doodson 
contant \citep[as defined by][]{2004A&A...427.1075P}.
The consequence could be that the planet may get lost in the star in a time comparable to the age 
of the system if Q$_{*}$/k$_{2*}$ is less than 10$^7$, where Q$_{*}$ is the stellar  tidal energy dissipation 
factor and k$_{2*}$ is the stellar Love number. This preliminary analysis will be developed  in a 
forthcoming paper.

\section{Conclusions}

After the  discovery by the CoRoT satellite of  transit-like photometric events on the star \corotstar, a planet that we called  \corotplanet\,  was eventually confirmed using ground-based photometric and spectroscopic follow-up observations. 
The amplitude of transits is $\Delta F / F \approx 4.3\, 10^{-3}$ $\pm$ 0.1 10$^{-3}$, as detected by the 
satellite. The star, characterized with high-resolution spectroscopy  has 
the spectral type  G0 V and is considered to be mature, i.e.  close to leaving  the main sequence.  The planetary mass resulting from RV measurements is \mplanet\, $\approx$  2.8 \MJ. The planetary orbital period,  3.6314 days, indicates that the planet
belongs to the now classical  {\it hot Jupiter} class.  

What is less classical is  that it features an eccentricity at the significant  level of 0.16. A second intriguing pecularity is the density  $\rho_{p}$ =   3 $g\;cm^{-3}$, which makes \corotplanet\,  among  the densest exoplanets of this  category.  Those two properties clearly do not usually pertain to that class of  hot Jupiters. 

We, however, show that both  characteristics are not all that extraordinary, because the density is likely the consequence of the long  duration of the planet  contraction and maybe of a rather massive core, while the eccentricity is consistent with a tide-driven damping during 5-10 Gyr of  a medium mass Jupiter at an orbital distance on the order of 0.05 AU. \corotplanet\, very likely belongs  to this class, recently identified, of exoplanets  that are  within the zone of the parameters space where a complete circularization is not achieved, even after 7.5 Gyr.

\begin{acknowledgements}
The authors are grateful to all the people that have worked on and operated 
the \corot\, satellite. The team at the IAC acknowledges support by grants
ESP2007-65480-C02-02 and AYA2010-20982-C02-02 of the Spanish Ministry
of Science and Innovation (MICINN).  The CoRoT/Exoplanet catalogue
(Exo-Dat) was made possible by observations collected for years at the
Isaac Newton Telescope (INT), operated on the island of La Palma by
the Isaac Newton group in the Spanish Observatorio del Roque de Los
Muchachos of the Instituto de Astrophysica de Canarias.
The German CoRoT Team (Th\"{u}ringer Landessternwarte and
       University of Cologne) acknowledges the support of grants
       50OW0204, 50OW603, and 50QM1004 from the Deutsches Zentrum f
       \"{u}r Luft- und Raumfahrt e.V. (DLR).
\end{acknowledgements}

\bibliographystyle{aa}
\bibliography{corotbib}

\begin{thebibliography}{36}
\expandafter\ifx\csname natexlab\endcsname\relax\def\natexlab#1{#1}\fi

\bibitem[{{Baglin} {et~al.}(2006){Baglin}, {Auvergne}, {Boisnard}, {Lam-Trong},
  {Barge}, {Catala}, {Deleuil}, {Michel}, \& {Weiss}}]{2006cosp...36.3749B}
{Baglin}, A., {Auvergne}, M., {Boisnard}, L., {et~al.} 2006, in COSPAR, Plenary
  Meeting, Vol.~36, 36th COSPAR Scientific Assembly, 3749

\bibitem[{{Baraffe} {et~al.}(2008){Baraffe}, {Chabrier}, \&
  {Barman}}]{2008A&A...482..315B}
{Baraffe}, I., {Chabrier}, G., \& {Barman}, T. 2008, \aap, 482, 315

\bibitem[{{Baranne} {et~al.}(1996){Baranne}, {Queloz}, {Mayor}, {Adrianzyk},
  {Knispel}, {Kohler}, {Lacroix}, {Meunier}, {Rimbaud}, \&
  {Vin}}]{1996A&AS..119..373B}
{Baranne}, A., {Queloz}, D., {Mayor}, M., {et~al.} 1996, \aaps, 119, 373

\bibitem[{{Boulade} {et~al.}(2003){Boulade}, {Charlot}, {Abbon}, {Aune},
  {Borgeaud}, {Carton}, {Carty}, {Da Costa}, {Deschamps}, {Desforge},
  {Eppell{\'e}}, {Gallais}, {Gosset}, {Granelli}, {Gros}, {de Kat}, {Loiseau},
  {Ritou}, {Rouss{\'e}}, {Starzynski}, {Vignal}, \&
  {Vigroux}}]{2003SPIE.4841...72B}
{Boulade}, O., {Charlot}, X., {Abbon}, P., {et~al.} 2003, in Society of
  Photo-Optical Instrumentation Engineers (SPIE) Conference Series, Vol. 4841,
  Society of Photo-Optical Instrumentation Engineers (SPIE) Conference Series,
  ed. {M.~Iye \& A.~F.~M.~Moorwood}, 72--81

\bibitem[{{Carpano} {et~al.}(2009){Carpano}, {Cabrera}, {Alonso}, {Barge},
  {Aigrain}, {Almenara}, {Bord{\'e}}, {Bouchy}, {Carone}, {Deeg}, {de La Reza},
  {Deleuil}, {Dvorak}, {Erikson}, {Fressin}, {Fridlund}, {Gondoin}, {Guillot},
  {Hatzes}, {Jorda}, {Lammer}, {L{\'e}ger}, {Llebaria}, {Magain}, {Moutou},
  {Ofir}, {Ollivier}, {Janot-Pacheco}, {P{\"a}tzold}, {Pont}, {Queloz},
  {Rauer}, {R{\'e}gulo}, {Renner}, {Rouan}, {Samuel}, {Schneider}, \&
  {Wuchterl}}]{2009A&A...506..491C}
{Carpano}, S., {Cabrera}, J., {Alonso}, R., {et~al.} 2009, \aap, 506, 491

\bibitem[{{Carpenter}(2001)}]{2001AJ....121.2851C}
{Carpenter}, J.~M. 2001, \aj, 121, 2851

\bibitem[{{Charbonneau} {et~al.}(2000){Charbonneau}, {Brown}, {Latham}, \&
  {Mayor}}]{2000ApJ...529L..45C}
{Charbonneau}, D., {Brown}, T.~M., {Latham}, D.~W., \& {Mayor}, M. 2000, \apjl,
  529, L45

\bibitem[{{Deeg} {et~al.}(2009){Deeg}, {Gillon}, {Shporer}, {Rouan},
  {Stecklum}, {Aigrain}, {Alapini}, {Almenara}, {Alonso}, {Barbieri}, {Bouchy},
  {Eisl{\"o}ffel}, {Erikson}, {Fridlund}, {Eigm{\"u}ller}, {Handler}, {Hatzes},
  {Kabath}, {Lendl}, {Mazeh}, {Moutou}, {Queloz}, {Rauer}, {Rabus}, {Tingley},
  \& {Titz}}]{2009A&A...506..343D}
{Deeg}, H.~J., {Gillon}, M., {Shporer}, A., {et~al.} 2009, \aap, 506, 343

\bibitem[{{Deeg} {et~al.}(2010){Deeg}, {Moutou}, {Erikson}, {Csizmadia},
  {Tingley}, {Barge}, {Bruntt}, {Havel}, {Aigrain}, {Almenara}, {Alonso},
  {Auvergne}, {Baglin}, {Barbieri}, {Benz}, {Bonomo}, {Bord{\'e}}, {Bouchy},
  {Cabrera}, {Carone}, {Carpano}, {Ciardi}, {Deleuil}, {Dvorak},
  {Ferraz-Mello}, {Fridlund}, {Gandolfi}, {Gazzano}, {Gillon}, {Gondoin},
  {Guenther}, {Guillot}, {Hartog}, {Hatzes}, {Hidas}, {H{\'e}brard}, {Jorda},
  {Kabath}, {Lammer}, {L{\'e}ger}, {Lister}, {Llebaria}, {Lovis}, {Mayor},
  {Mazeh}, {Ollivier}, {P{\"a}tzold}, {Pepe}, {Pont}, {Queloz}, {Rabus},
  {Rauer}, {Rouan}, {Samuel}, {Schneider}, {Shporer}, {Stecklum}, {Street},
  {Udry}, {Weingrill}, \& {Wuchterl}}]{2010Natur.464..384D}
{Deeg}, H.~J., {Moutou}, C., {Erikson}, A., {et~al.} 2010, \nat, 464, 384

\bibitem[{{Deleuil} {et~al.}(2011){Deleuil}, {Bonomo}, {Ferraz-Mello},
  {Erikson}, {Bouchy}, \& {Guillot}}]{2011A&A..Deleuil}
{Deleuil}, M., {Bonomo}, A., {Ferraz-Mello}, S., {et~al.} 2011, \aap, submitted

\bibitem[{{Deleuil} {et~al.}(2009){Deleuil}, {Meunier}, {Moutou}, {Surace},
  {Deeg}, {Barbieri}, {Debosscher}, {Almenara}, {Agneray}, {Granet},
  {Guterman}, \& {Hodgkin}}]{2009AJ....138..649D}
{Deleuil}, M., {Meunier}, J.~C., {Moutou}, C., {et~al.} 2009, \aj, 138, 649

\bibitem[{{Ducati} {et~al.}(2001){Ducati}, {Bevilacqua}, {Rembold}, \&
  {Ribeiro}}]{2001ApJ...558..309D}
{Ducati}, J.~R., {Bevilacqua}, C.~M., {Rembold}, S.~B., \& {Ribeiro}, D. 2001,
  \apj, 558, 309

\bibitem[{{Dutra} {et~al.}(2003){Dutra}, {Santiago}, {Bica}, \&
  {Barbuy}}]{2003MNRAS.338..253D}
{Dutra}, C.~M., {Santiago}, B.~X., {Bica}, E.~L.~D., \& {Barbuy}, B. 2003,
  \mnras, 338, 253

\bibitem[{{Frasca} {et~al.}(2003){Frasca}, {Alcal{\'a}}, {Covino}, {Catalano},
  {Marilli}, \& {Paladino}}]{2003A&A...405..149F}
{Frasca}, A., {Alcal{\'a}}, J.~M., {Covino}, E., {et~al.} 2003, \aap, 405, 149

\bibitem[{{Gandolfi} {et~al.}(2008){Gandolfi}, {Alcal{\'a}}, {Leccia},
  {Frasca}, {Spezzi}, {Covino}, {Testi}, {Marilli}, \&
  {Kainulainen}}]{2008ApJ...687.1303G}
{Gandolfi}, D., {Alcal{\'a}}, J.~M., {Leccia}, S., {et~al.} 2008, \apj, 687,
  1303

\bibitem[{{Gregory}(2007)}]{2007MNRAS.381.1607G}
{Gregory}, P.~C. 2007, \mnras, 381, 1607

\bibitem[{{Guillot}(2008)}]{2008PhST..130a4023G}
{Guillot}, T. 2008, Physica Scripta Volume T, 130, 014023

\bibitem[{{Guillot} \& {Havel}(2011)}]{2011A&A...527A..20G}
{Guillot}, T. \& {Havel}, M. 2011, \aap, 527, A20+

\bibitem[{{Guillot} \& {Morel}(1995)}]{1995A&AS..109..109G}
{Guillot}, T. \& {Morel}, P. 1995, \aaps, 109, 109

\bibitem[{{Havel} {et~al.}(2011){Havel}, {Guillot}, {Valencia}, \&
  {Crida}}]{2011A&A...531A...3H}
{Havel}, M., {Guillot}, T., {Valencia}, D., \& {Crida}, A. 2011, \aap, 531, A3+

\bibitem[{{Hebb} {et~al.}(2009){Hebb}, {Collier-Cameron}, {Loeillet},
  {Pollacco}, {H{\'e}brard}, {Street}, {Bouchy}, {Stempels}, {Moutou},
  {Simpson}, {Udry}, {Joshi}, {West}, {Skillen}, {Wilson}, {McDonald},
  {Gibson}, {Aigrain}, {Anderson}, {Benn}, {Christian}, {Enoch}, {Haswell},
  {Hellier}, {Horne}, {Irwin}, {Lister}, {Maxted}, {Mayor}, {Norton}, {Parley},
  {Pont}, {Queloz}, {Smalley}, \& {Wheatley}}]{2009ApJ...693.1920H}
{Hebb}, L., {Collier-Cameron}, A., {Loeillet}, B., {et~al.} 2009, \apj, 693,
  1920

\bibitem[{{Hebb} {et~al.}(2010){Hebb}, {Collier-Cameron}, {Triaud}, {Lister},
  {Smalley}, {Maxted}, {Hellier}, {Anderson}, {Pollacco}, {Gillon}, {Queloz},
  {West}, {Bentley}, {Enoch}, {Haswell}, {Horne}, {Mayor}, {Pepe}, {Segransan},
  {Skillen}, {Udry}, \& {Wheatley}}]{2010ApJ...708..224H}
{Hebb}, L., {Collier-Cameron}, A., {Triaud}, A.~H.~M.~J., {et~al.} 2010, \apj,
  708, 224

\bibitem[{{Hellier} {et~al.}(2010){Hellier}, {Anderson}, {Collier Cameron},
  {Gillon}, {Lendl}, {Maxted}, {Queloz}, {Smalley}, {Triaud}, {West}, {Brown},
  {Enoch}, {Lister}, {Pepe}, {Pollacco}, {S{\'e}gransan}, \&
  {Udry}}]{2010ApJ...723L..60H}
{Hellier}, C., {Anderson}, D.~R., {Collier Cameron}, A., {et~al.} 2010, \apjl,
  723, L60

\bibitem[{{Kipping}(2010)}]{2010MNRAS.407..301K}
{Kipping}, D.~M. 2010, \mnras, 407, 301

\bibitem[{{Marigo} {et~al.}(2008){Marigo}, {Girardi}, {Bressan}, {Groenewegen},
  {Silva}, \& {Granato}}]{2008A&A...482..883M}
{Marigo}, P., {Girardi}, L., {Bressan}, A., {et~al.} 2008, \aap, 482, 883

\bibitem[{{Matsumura} {et~al.}(2010){Matsumura}, {Peale}, \&
  {Rasio}}]{2010ApJ...725.1995M}
{Matsumura}, S., {Peale}, S.~J., \& {Rasio}, F.~A. 2010, \apj, 725, 1995

\bibitem[{{Moeckel} {et~al.}(2008){Moeckel}, {Raymond}, \&
  {Armitage}}]{2008DDA....39.1515M}
{Moeckel}, N., {Raymond}, S.~N., \& {Armitage}, P.~J. 2008, in AAS/Division of
  Dynamical Astronomy Meeting, Vol.~39, AAS/Division of Dynamical Astronomy
  Meeting 39, 15.15--+

\bibitem[{{Morel} \& {Lebreton}(2008)}]{2008Ap&SS.316...61M}
{Morel}, P. \& {Lebreton}, Y. 2008, \apss, 316, 61

\bibitem[{{P{\"a}tzold} {et~al.}(2004){P{\"a}tzold}, {Carone}, \&
  {Rauer}}]{2004A&A...427.1075P}
{P{\"a}tzold}, M., {Carone}, L., \& {Rauer}, H. 2004, \aap, 427, 1075

\bibitem[{{P{\"a}tzold} \& {Rauer}(2002)}]{2002ApJ...568L.117P}
{P{\"a}tzold}, M. \& {Rauer}, H. 2002, \apjl, 568, L117

\bibitem[{{Pepe} {et~al.}(2002){Pepe}, {Mayor}, {Galland}, {Naef}, {Queloz},
  {Santos}, {Udry}, \& {Burnet}}]{2002A&A...388..632P}
{Pepe}, F., {Mayor}, M., {Galland}, F., {et~al.} 2002, \aap, 388, 632

\bibitem[{{Pont} {et~al.}(2011){Pont}, {Husnoo}, {Mazeh}, \&
  {Fabrycky}}]{2011MNRAS.414.1278P}
{Pont}, F., {Husnoo}, N., {Mazeh}, T., \& {Fabrycky}, D. 2011, \mnras, 414,
  1278

\bibitem[{{Rouan} {et~al.}(1998){Rouan}, {Baglin}, {Copet}, {Schneider},
  {Barge}, {Deleuil}, {Vuillemin}, \& {L{\'e}ger}}]{1998EM&P...81...79R}
{Rouan}, D., {Baglin}, A., {Copet}, E., {et~al.} 1998, Earth Moon and Planets,
  81, 79

\bibitem[{{Straizys} \& {Kuriliene}(1981)}]{1981Ap&SS..80..353S}
{Straizys}, V. \& {Kuriliene}, G. 1981, \apss, 80, 353

\bibitem[{{Valenti} \& {Piskunov}(1996)}]{1996A&AS..118..595V}
{Valenti}, J.~A. \& {Piskunov}, N. 1996, \aaps, 118, 595

\bibitem[{{Veras} \& {Ford}(2009)}]{2009ApJ...690L...1V}
{Veras}, D. \& {Ford}, E.~B. 2009, \apjl, 690, L1

\end{thebibliography}

\end{document}